\documentclass[aps,twocolumn,superscriptaddress]{revtex4}
\usepackage{amsmath}
\usepackage{amsfonts}
\usepackage{mathrsfs}
\usepackage{graphicx}
\usepackage{times}
\usepackage{color}

\def\be{\begin{equation}}
\def\ee{\end{equation}}
\def\bea{\begin{eqnarray}}
\def\eea{\end{eqnarray}}

\begin{document}

\title{Orbital hybridized topological Fulde-Ferrel superfluidity in a noncentrosymmetric optical lattice}

\author{Bo Liu}
\affiliation{Department of Physics and Astronomy, University of
Pittsburgh, Pittsburgh, PA 15260, USA}
\affiliation{Wilczek Quantum
Center, Zhejiang University of Technology, Hangzhou 310023, China}
\author{Xiaopeng Li}
\affiliation{Condensed Matter Theory Center and Joint Quantum
Institute, University of Maryland, College Park, MD 20742, USA}
\author{W. Vincent Liu}
\affiliation{Department of Physics and Astronomy, University of
Pittsburgh, Pittsburgh, PA 15260, USA}
\affiliation{Wilczek Quantum
Center, Zhejiang University of Technology, Hangzhou 310023, China}
\date{\today}

\begin{abstract}

Topological phases like topological insulators or superconductors
are fascinating quantum states of matter, featuring novel properties
such as emergent chiral edge states or Majorana fermions with
non-Abelian braiding statistics. The recent experimental
implementation of optical lattices with highly tunable geometry in
cold gases opens up a new thrust on exploring these novel quantum
states. {Here we report that the topological non-trivial Bloch bands
can arise naturally in a noncentrosymmetric lattice. It induces a
controllable orbital hybridization, producing the topological band
structure. In such bands, when considering attractive fermionic
atoms, we find a topological Fulde-Ferrell superfluid state with
finite center-of-mass momentum in the presence of onsite rotation.}
This topological superfluid supports Majorana fermions on its edges.
Experimental signatures are predicted for cold gases in
radio-frequency spectroscopy.

\end{abstract}

\maketitle

Symmetry plays an important role in solid state materials and
influences many of their properties in a profound way. Recently, the
noncentrosymmetric materials, i.e., crystal structure lacking a
center of inversion, have attracted considerable theoretical and
experimental attention in various fields of condensed matter
physics~\cite{2009_Samokhin_AOP}. In contrast to the centrosymmetric
case, the absence of inversion symmetry leads to {very rich physical
phenomena}, such as skyrmion
states~\cite{2009_Mulbauer_science,2010_Yu_nature,2012_Seki_science},
novel superconducting
phases~\cite{2004_SamokhinPhysRevB,2008_Mineev_PhysRevB,2011a_Schnyder_PhysRevB,2011b_Schnyder_PhysRevB},
as well as magnetoelectric
effect~\cite{2005_Fujimoto_PhysRevB,2002_Yip_PhysRevB}. In parallel
to the developments in solid state physics, optical lattices with
highly tunable geometry in the recent ultracold atom
experiments~\cite{2012_Esslinger_Nature} provide new opportunities
to study noncentrosymmetric materials. {Taking the advantage of high
controllability, the cold gases are expected to emerge {not only as
a flexible tool to simulate electronic system, but also a versatile
platform to create fascinating quantum states of matter not existing
in nature.}}

{Recent experimental progress in manipulating higher orbital bands
in optical
lattices~\cite{2007_Bloch_PRL,2011_Hemmerich_NatPhys,2012_Sengstock_Natphys,2013_cheng_Natphys,2015_Hemmerich_PhysRevLett}
provides unprecedented  opportunities to investigate quantum
many-body phases with orbital degrees of freedom. Studying higher
orbital physics in optical lattices is attracting considerable
interests due to their unique and intrinsic spatial
nature~\cite{2011_Vincent_naturephy}. An idea of using orbital
hybridization to emulate spin-orbit physics, or artificial gauge
fields in general, has  emerged in recent theoretical
studies~\cite{2012_Kaisun_naturephy,2013_xiaopeng_Natcommun,2014_Bo_arxiv,2008_congjun_PhysRevLett,
2008_2_congjun_PhysRevLett}, where various interesting many-body
states have been proposed. In the context of periodically driven
systems, orbital hybridization is shown to be controllable by
lattice
shaking~\cite{2014_Qizhou_PhysRevA,2014_huizhai_PhysRevA,2014_cooper_PhysRevA}.
However how to systematically control the orbital hybridization in
static optical lattices remains unclear and stands as an obstacle to
explore the rich phenomena in orbital hybridized many-body ground
states.}

{In this letter, we propose a static noncentrosymmetric optical
lattice, where the orbital hybridization is systematically
controllable by manipulating symmetry breaking. We find topological
bands arise from the interplay between high orbitals and inversion
symmetry breaking, yet without requiring Raman induced spin-orbit
coupling nor other artificial gauge
field~\cite{2013_Galitski_nature,2012_huizhai_IJMP}. Furthermore
when considering attractive fermionic atoms, say
$^6$Li~\cite{2015_Hart_nature,2010_Hulet_nature,2006_Hulet,2006_Martin_Science}
loaded into such a noncentrosymmetric optical lattice, we find an
orbital hybridized topological Fulde-Ferrell superfluid state (tFF)
in the presence of local orbital angular momentum induced by onsite
rotation~\cite{2010_StevenChu_arxiv}. This state represents an
unconventional type of Fulde-Ferrell-Larkin-Ovchinnikov (FFLO)
state, topologically distinct from the $s$-wave FFLO superfluidity
observed in a quasi-one dimensional Fermi
gas~\cite{2010_Hulet_nature}. This orbital realization of the tFF
superfluid state here is distinguished from previous works based on
hyperfine
states~\cite{2013_Qu_Natcommun,2013_Zhang_Natcommun,2015_xuyong_IJMP},
where spin-orbit coupling and Zeeman fields are required. Moreover,
our symmetry-based method of controlling orbital hybridization is in
principle applicable to other optical lattice setups as
well~\cite{2012_Kaisun_naturephy,2013_xiaopeng_Natcommun,2014_Bo_arxiv,
2012_xiaopeng_PhysRevLett,2008_erhai_PhysRevLett,2008_congjun_PhysRevLett,2008_2_congjun_PhysRevLett},
which could lead to more interesting noncentrosymmetric many-body
phases worth future exploring.}

\begin{figure}[t]
\begin{center}
\includegraphics[width=9cm]{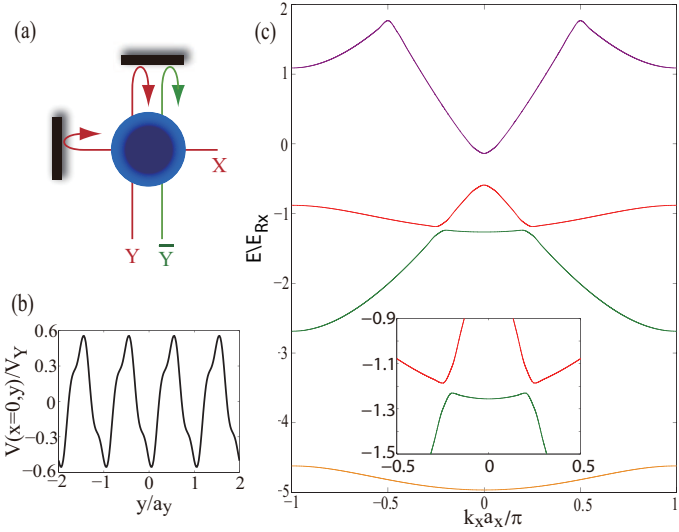}
\end{center}
\caption{{Noncentrosymmetric optical lattice and topological band
structures.} (a) Three retro-reflected laser beams create the
lattice potential in Eq.~\eqref{Lattice}. $X$ and $Y$ interfere and
produce a 2D square lattice, while $\bar{Y}$ creates an independent
standing wave. (b) Lattice potential along the $y$ direction shows
the inversion symmetry breaking. (c) The single-particle energy
spectrum along $k_x$ axis in the unit of $E_{Rx}$ for the lowest
four bands through plane wave expansion calculation. Here the
lattice depth is chosen as ${V_X/E_{Rx}}=4$, ${V_Y/E_{Ry}}=40$ and
${V_{\bar{Y}}/E_{Ry}}=4$ with the recoil energy $E_{Rx}={\hbar^2
k^2_{Lx}}/{2m}$ and $E_{Ry}={\hbar^2 k^2_{Ly}}/{2m}$.}
\label{fig:lattice}
\end{figure}

\textit{Effective model.} Let us consider a noncentrosymmetric
optical lattice with the potential
\begin{eqnarray}
V(x,y)&=&-V_X \cos^2(k_{Lx}x)-V_Y \cos^2(k_{Ly}y)\notag \\
&+&V_{\bar{Y}} \cos^2(3{k}_{Ly}y+\theta/2), \label{Lattice}
\end{eqnarray}
where $V_X$, $V_Y$, $V_{\bar{Y}}$ are the depth of optical lattices,
$k_{Lx}$, $k_{Ly}$ are the wavevectors of laser fields and the
corresponding lattice constants are $a_x=\pi/k_{Lx}$,
$a_y=\pi/k_{Ly}$ along $x$ and $y$ directions respectively. By the
techniques of designing the geometry of optical lattices developed
in the recent experimental advances~\cite{2012_Esslinger_Nature},
the configuration of optical lattices considered here can be formed
through three retro-reflected laser beams as shown in
Fig.~\ref{fig:lattice}(a). The interference of two perpendicular
beams $X$ and $Y$ gives rise to a 2D square lattice. {The} third
beam $\bar{Y}$ creates an additional standing wave pattern which
breaks the inversion symmetry along the $y$ direction, for example,
when $\theta=\pi/2$ as shown in Fig.~\ref{fig:lattice}(b). This
noncentrosymmetric geometry plays a crucial role in producing the
non-trivial Bloch bands in our model, to be illustrated below. Here
we consider highly anisotropic geometry of the optical lattice
through increasing lattice depth ($V_Y \gg V_X$) and spacing ($a_y
\gg a_x$) in the $y$ direction, meanwhile requiring that the lattice
potential satisfy the condition $V_Xk^2_{Lx}=V_Yk^2_{Ly}$.  This
should keep local rotational symmetry of each site in the $xy$
plane~\cite{2012_xiaopeng_PhysRevLett}. {The band structure of such
a lattice system is solved numerically through plane wave expansion.
We find that the second and third bands cross in the absence of
$V_{\bar{Y}}$ with the band touching points protected by
time-reversal and inversion symmetries. With finite $V_{\bar{Y}}$,
we see the gap reopening due to inversion symmetry breaking
(Fig.~\ref{fig:lattice}(c)). The band mixing can thus be turned on
and off by controlling the symmetry of the lattice geometry.}

{The essential physics of band mixing is captured by the following
multi-orbital tight binding model without considering the full band
structure theory,}
\begin{eqnarray}
H_0&=&t_{x}\sum_{\mathbf{r}}C_{p_x}^{\dagger
}(\mathbf{r})C_{p_x}(\mathbf{r}+\vec{e}_{x})-t_{y}\sum_{\mathbf{r}}C_{p_y}^{\dagger
}(\mathbf{r})C_{p_y}(\mathbf{r}+\vec{e}
_{x}) \notag \\
&+&t\sum_{\mathbf{r}}C_{p_x}^{\dagger }(\mathbf{r})C_{p_y}(\mathbf{r}+\vec{e}_{x})-t \sum_{\mathbf{r}} C_{p_y}^{\dagger }
(\mathbf{r})C_{p_x}(\mathbf{r}+\vec{e}_{x}) \notag \\
&-&t^{\prime}_{x}\sum_{\mathbf{r}}C_{p_x}^{\dagger
}(\mathbf{r})C_{p_x}(\mathbf{r}+\vec{e}_{y})+t^{\prime}_{y}\sum_{\mathbf{r}}
C_{p_y}^{\dagger}(\mathbf{r})C_{p_y}(\mathbf{r}+\vec{e}_{y}) \notag \\
&+&h.c.-\mu \sum_{\mathbf{r}}[C_{p_x}^{\dagger
}(\mathbf{r})C_{p_x}(\mathbf{r})+C_{p_y}^{\dagger }
(\mathbf{r})C_{p_y}(\mathbf{r})],
\label{FreeHam}
\end{eqnarray}
where $C_{\nu }(\mathbf{r})$ is a fermionic annihilation operator
for the localized $\nu$ orbital ($p_{x}$ or $p_{y}$) located at the
lattice site $\mathbf{r}$ and the chemical potential is denoted by
$\mu$. Since the system is relatively stronger confined in the $y$
direction, the tunnelings $(t^{\prime}_{x},t^{\prime}_{y})$ in the
$y$ direction are much weaker compared to that ($t_{x},t_{y}$) along
the $x$ direction. The relative sign of the hopping amplitude is
fixed by the parity of $p_x$ and $p_y$ orbitals. {The key ingredient
in our model is the hybridization between $p_x$ and $p_y$ orbitals.
It arises from the asymmetric shape of the $p_y$ orbital
{wavefunction} induced by the inversion symmetry breaking in the $y$
direction. {This asymmetry leads} to the orbital hybridization
$t\sum_{\mathbf{r}}[C_{p_x}^{\dagger
}(\mathbf{r})C_{p_y}(\mathbf{r}+\vec{e}_{x})-C_{p_y}^{\dagger
}(\mathbf{r})C_{p_x}(\mathbf{r}+\vec{e}_{x})]+h.c.$ in
Eq.~\eqref{FreeHam}, which plays a crucial role in producing
topological non-trivial band structures.} It resembles spin-orbit
coupling~\cite{2011_Lin_nature,2012_Cheuk_PhysRevLett,2012_Wang_PhysRevLett}
when the $p_x$ and $p_y$ orbitals are mapped to pseudo-spin-$1/2$
states. {But unlike the spin-orbit coupling, the orbital
hybridization is geometrically controllable through manipulating the
inversion symmetry of the optical lattice.}

\begin{figure}[t]
\begin{center}
\includegraphics[width=9cm]{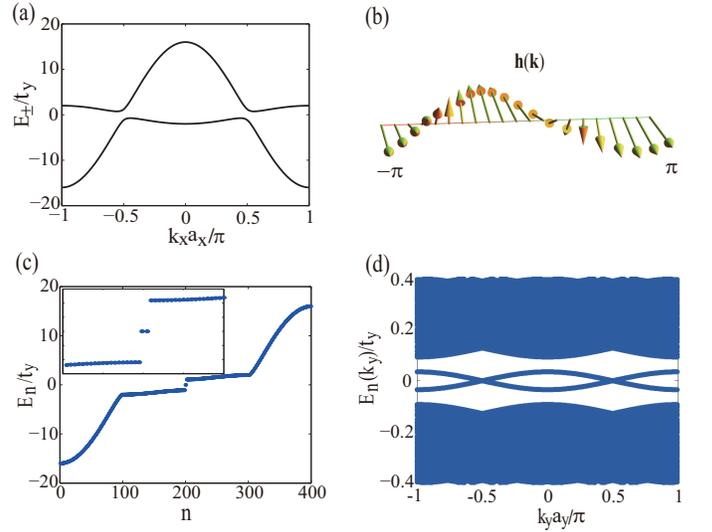}
\end{center}
\caption{{Bulk and edge properties of the topological band
insulator.} (a) The single-particle band structure calculated from
the tight binding model (Eq.~\eqref{FreeHam}) in quasi-1D limit,
when $t_x/t_y=8$, $t/t_y=0.6$. (b) Topological winding of
Hamiltonian in Eq.~\eqref{FreeHamMS} across the Brillouin zone. (c)
{Energy spectrum of the Hamiltonian (Eq.~\eqref{FreeHam}) in
quasi-1D limit with finite length (lattice sites $N=200$),} when
$t_x/t_y=8$ and $t/t_y=0.9$. There are {two} zero-energy states
inside the gap which are located at the two outer edges of the
system respectively. {Here $n$ labels the energy level.} (d) The
effect of the small transverse tunneling. The midgap bands show a
finite curvature along the $y$ direction with open (and periodic)
boundary conditions in the $x$ ($y$) directions when $t_x/t_y=8$,
$t/t_y=0.1$, $t^\prime_y/t_y=0.02$ and $t^\prime_x/t^\prime_y=0.1$.}
\label{fig:TIedge}
\end{figure}

\textit{Topological band structures.} {We first focus on the
quasi-one dimensional limit where the transverse ($y$ direction)
tunnelings are negligible.} {At half filling (one particle per
lattice site), in the basis of $(C_{p_x}^{\dagger
}(\mathbf{k}),C_{p_y}^{\dagger}(\mathbf{k}))$, the Hamiltonian takes
a suggestive form
\begin{equation}
H_0(\mathbf{k})=h_0(\mathbf{k})\mathbb{I}+\mathbf{h}(\mathbf{k})\cdot
{\mathbf {\sigma}}, \label{FreeHamMS}
\end{equation}
where $h_0(\mathbf{k})=(t_x-t_y)\cos(k_xa_x)$, $h_x(\mathbf{k})=0$,
$h_y(\mathbf{k})=-2t\sin(k_xa_x)$,
$h_z(\mathbf{k})=(t_x+t_y)\cos(k_xa_x)$, and $\sigma_{x,y,z}$ are
Pauli matrices.} The energy spectrum reads
$E_{\pm}=h_0(\mathbf{k})\pm\sqrt{h^2_y(\mathbf {k})+h^2_z(\mathbf
{k})}$. As shown in Fig.~\ref{fig:TIedge}(a), {the system is in an
insulating state with a band gap determined by the orbital
hybridization $t$.} It is a $Z_2$ topological insulator protected by
time reversal and reflection
symmetries~\cite{2008_Schnyder_PhysRevB}. To visualize the
topological properties of the band structures, we show the vector
$\mathbf{h}(\mathbf{k})$ winds an angle of $2\pi$ when the momentum
$\mathbf{k}$ varies from $-\pi$ to $\pi$ crossing the entire
Brillouin zone (BZ) in Fig.~\ref{fig:TIedge}(b). It is also
confirmed from the calculation of winding number defined as $W=\oint
\frac{dk_x}{4\pi}\epsilon_{\eta \eta'}
\hat{h}^{-1}_{\eta}(k_x)\partial_{k_x}\hat{h}_{\eta}$ with $\hat{h}
\equiv \frac{\mathbf{h}}{|\mathbf {h}|}$ and
$\epsilon_{yz}=-\epsilon_{zy}=1$. When $t\neq 0$, the winding number
is $1$, which signifies a topological band insulator state. The
non-trivial topology of this state also manifests through the
existence of the edge states. As shown in Fig.~\ref{fig:TIedge}(c),
there are two emergent zero-energy modes located at the two outer
edges of the system respectively.

Next we will discuss the effect of the small transverse ($y$
direction) tunneling, which has been neglected above but always
exists in a realistic quasi-one dimensional system. {By considering
small transverse tunneling, the zero-energy modes of individual
chain will morph into a midgap band, with finite curvature in the
transverse direction as shown in Fig.~\ref{fig:TIedge}(d).} The
topological band insulator state remains stable at small value of
transverse tunneling. For example, when $t_x/t_y=8$ and $t/t_y=0.1$,
the topological state survives until $t^{\prime}_y/t_y$ reaches
$0.073$ with $t^{\prime}_x=0.1t^{\prime}_y$. However, beyond this
value, the band gap will close and the topological band insulator
state becomes unstable.

\textit{Topological Fulde-Ferrell state and Majorana fermions.} In
this section, {we study attractive fermions in the topological
noncentrosymmetric optical lattice (Eq.~\eqref{Lattice}) with onsite rotation~\cite{2010_StevenChu_arxiv} and show
that a topological Fulde-Ferrell superfluid state (tFF) with finite
center-of-mass momentum emerges. The interacting model to describe
this fermionic system is}
\begin{equation}
H=H_0+H_{int}+H_{L}+H_{Z}. \label{Ham}
\end{equation}
{In this model, the interaction $H_{int}=U\sum_{{\mathbf
{r}}}C_{p_x}^{\dagger
}(\mathbf{r})C_{p_x}(\mathbf{r})C_{p_y}^{\dagger
}(\mathbf{r})C_{p_y}(\mathbf{r})$ can be induced by considering
optical Feshbach
resonance~\cite{2013_Takahashi_PhysRevA,2010_Goyal_PhysRevA},
Bose-Fermi mixtures~\cite{2000_Bijlsma_PhysRevA} or dipolar
atoms$\setminus$molecules~\cite{2012_Baranov_Reviews}.
$H_{L}=i\Omega_z\sum_{{\mathbf {r}}}[C_{p_x}^{\dagger
}(\mathbf{r})C_{p_y}(\mathbf{r})-C_{p_y}^{\dagger
}(\mathbf{r})C_{p_x}(\mathbf{r})]$ is the onsite rotation induced
orbital Zeeman energy.  $H_{Z}$ is the orbital splitting $
h\sum_{{\mathbf {r}}}[C_{p_x}^{\dagger
}(\mathbf{r})C_{p_x}(\mathbf{r})-C_{p_y}^{\dagger
}(\mathbf{r})C_{p_y}(\mathbf{r})]$ originated from the onsite energy
difference between $p_x$ and $p_y$ orbitals, which is tunable
through adjusting the lattice depth or lattice spacing in $x$ and
$y$ directions.} {In the presence of onsite rotation, i.e.,
$\Omega_z \neq 0$, the Fermi surface of the system becomes
asymmetric along the $x$ direction. The pairing with the
center-of-mass momentum ${\mathbf {Q}}$ and ${\mathbf {-Q}}$ are no
longer degenerate. The Fulde-Ferrell (FF) state with a plane wave
order parameter $\propto e^{i\mathbf {Q}\cdot\mathbf {r}}$ is more
energetically favorable as compared to the Larkin-Ovchinnikov state
with a stripe order $\propto \cos({\bf Q \cdot {\bf r}})$.

{{Taking the superfluid pairing order parameter $\Delta(\mathbf
{r})=U\langle C_{p_y}(\mathbf {r})C_{p_x}(\mathbf {r})\rangle=\Delta
e^{i{\mathbf {Q}\cdot {\mathbf {r}}}}$, the system is described by
the Bogoliubov-de-Genes (BdG) Hamiltonian at mean field level.}
Since the weak transverse hopping introduces a small Fermi surface
curvature, we expect the center-of-mass momentum of pairing $\mathbf
{Q}$ pointing along the $x$ direction, say $\mathbf {Q}= Q(1, 0)$,
in order to maximize the phase space of pairing. Through
diagonalizing the BdG Hamiltonian, we obtain the free energy of the
system. The pairing order parameter $\Delta$ and the center-of-mass
momentum of pairing $\mathbf {Q}$ are determined from minimizing the
free energy. The details are given in the Supplemental Material.} We
find that when $\Omega_z > 0$ the center-of-mass momentum of pairing
along the $x$ direction is selected as a certain positive $Q$ due to
the deformation of the Fermi surface. {Since the $p_x$ and $p_y$
bands have different bandwidths, we introduce two dimensionless
quantities $\tilde{\mu}_{p_x}=\frac{\mu-h}{2t_x}$ and
$\tilde{\mu}_{p_y}=\frac{\mu+h}{2t_y}$, {which respectively control
the average filling of the two $p$ bands. Correspondingly, we also
introduce a dimensionless chemical potential
$\tilde{\mu}=\frac{\tilde{\mu}_{p_x}+\tilde{\mu}_{p_y}}{2}$ and
orbital polarization
$\tilde{h}=\frac{\tilde{\mu}_{p_y}-\tilde{\mu}_{p_x}}{2}$. For fixed
$\tilde{\mu}$, the resulting phase diagram as a function of
$\tilde{h}$ and the orbital hybridization $t$ is shown in
Fig.~\ref{fig:Phsaediagram}(a).} } There are two first order phase
transitions as the polarization $\tilde {h}$ is increased (except at
$t/t_y\approx 0.8$). {The first one is a transition from a gapped FF
superfluid to {a gapless FF superfluid state (with Bogoliubov
quasi-particles being gapless). The gap closing across the phase
transition is shown in Fig.~\ref{fig:Phsaediagram}(b).} } Further
increasing the polarization, the second one occurs between the
gapless FF and tFF superfluid states. When $t/t_y\approx 0.8$, there
is only one phase transition from FF to tFF superfluid states
without passing through the gapless FF superfluids, since we find in
Fig.~\ref{fig:Phsaediagram}(c) that $E_g$ will firstly close and
reopen immediately. We also find that a finite polarization is
required to stabilize the tFF superfluid state. The critical
polarization $\tilde{h}_{\rm c}$ decreases as the orbital
hybridization $t$ increases.

\begin{figure}[t]
\begin{center}
\includegraphics[width=9cm]{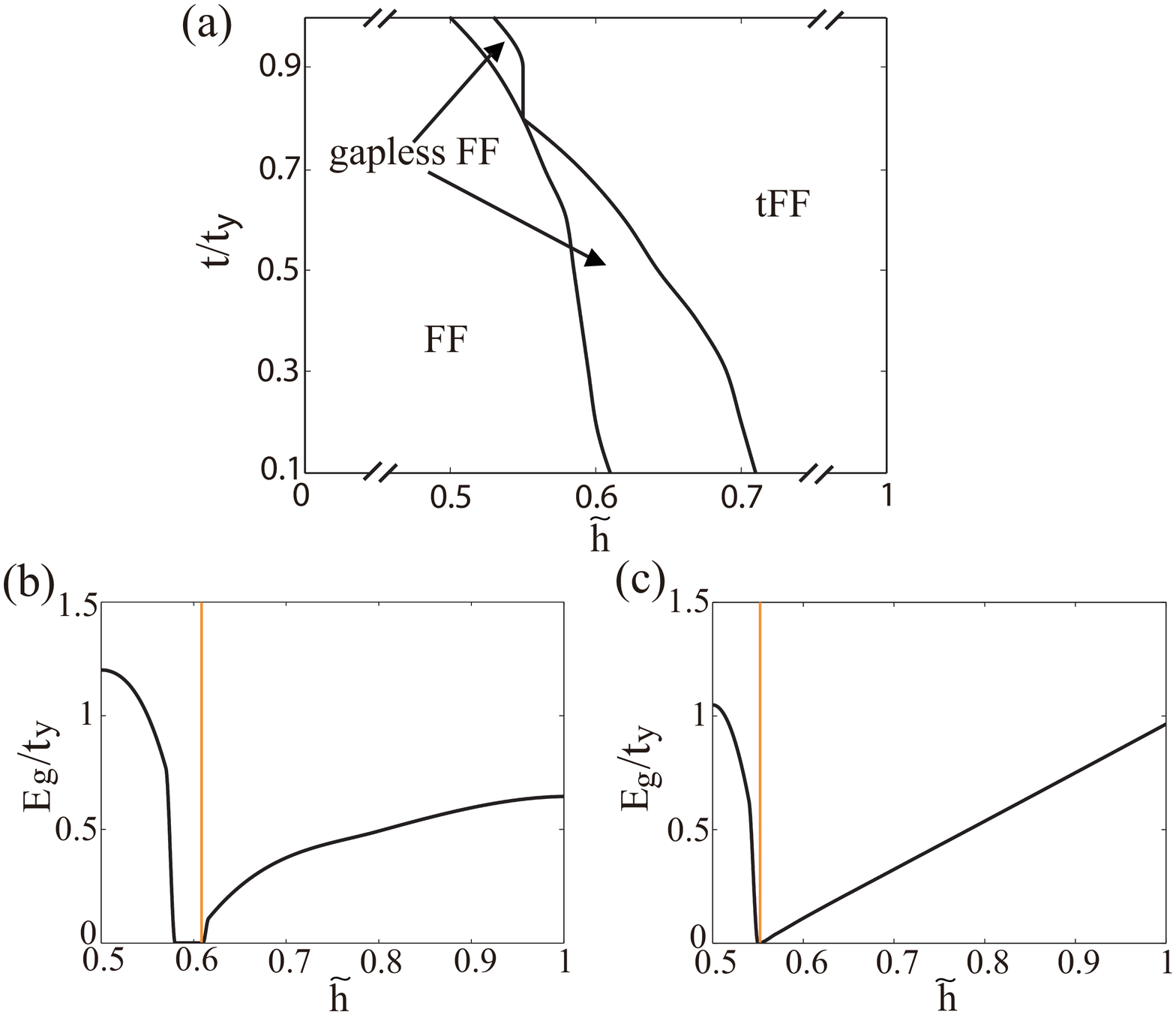}
\end{center}
\caption{{Zero-temperature phase diagram with fixed $\tilde{\mu}$.}
(a) When the polarization $\tilde{h}$ is small, the FF superfluid
state is the ground state of the system. By increasing $\tilde{h}$,
{a first order phase transition occurs between FF superfluidity and
a gapless FF superfluid state.} Further increasing $\tilde{h}$, the
system eventually evolves into topological FF superfluid state
(tFF). Other parameters are $t_x/t_y=8$, $t^\prime_y/t_y=0.05$,
$t^\prime_x/t^\prime_y=0.1$, $U/t_y=-15$, $\tilde{\mu}=0.4$, and
$\Omega_z/t_y=0.3$. (b) and (c) The quasi-particle exaction gap
$E_g$ as a function of polarization $\tilde{h}$ when $t/t_y=0.6$ and
$t/t_y=0.8$ respectively. The vertical lines mark the point where
the $Z_2$ topological invariant changes.} \label{fig:Phsaediagram}
\end{figure}

The transition from non-topological to topological FF states here
can be understood by observing the non-trivial $Z_2$ topological
invariant. {The BdG Hamiltonian (See details in the Supplemental
Material) maintains the particle-hole symmetry, i.e., $\Xi
H_{BdG}(\mathbf {k})\Xi^{-1}=-H_{BdG} ^{*} (\mathbf {-k})$, with
$\Xi=\begin{pmatrix} {0} & {\mathbb{I}}\\ {\mathbb{I}} & {0}
\end{pmatrix}$, while the time reversal and chiral symmetries are
broken. Therefore, the tFF superfluid state predicated here belongs
to the $D$ symmetry class according to the general classification
scheme of topological superconductors~\cite{2008_Schnyder_PhysRevB}.
{This topological state is thus characterized by a $Z_2$ topological
invariant~\cite{2001_Kitaev_phsUsp,2010_Tewair_PRB}. As shown in
Fig.~\ref{fig:Phsaediagram}(b) and (c), {we find that the $Z_2$
topological invariant $M=-1$ in the tFF superfluid state (see
details in the Supplemental Material).}

To further demonstrate the topological nature of the tFF superfluid
phase, we will show Majorana fermions {are} supported in this state.
To see this, we consider a cylinder geometry of the system, where
the open (periodic) boundary condition is chosen in the $x(y)$
direction respectively. {The energy spectrum in
Fig.~\ref{fig:tFFedge}(a)  is labeled by the momentum $k_y$. As
shown in Fig.~\ref{fig:tFFedge}(a), all the bulk modes are gapped
and there are two degenerate flat bands composed of Majorana
fermions located at the two outer edges of the system respectively.}
As shown in Fig.~\ref{fig:tFFedge}(c), for a fixed $k_y$, there are
two zero-energy states located at the two outer edges of the system.
{The corresponding wavefunctions $(u_{0,\nu},v_{0,\nu})^{T}$ satisfy
the relation $u_{0,\nu}(x)=v^*_{0,\nu}(x)$
[$u_{0,\nu}(x)=-v^*_{0,\nu}(x)$] on the right [left] edge
(Fig.~\ref{fig:tFFedge}(d) and (e)).} {These eigenstates {support}
localized Majorana fermions at the edges of the system. These
Majorana fermions are signified through the local density of states
(LDOS) which can be measured by radio-frequency (rf)
spectroscopy~\cite{2003_Ketterle_Science,2003_Jin_PRLrf,2007_Ketterle_PRL}.}
The LDOS is calculated as $\rho (x,E)=1/2\sum_{n,\nu}\int
dk_{y}[|u_{n,\nu }|^{2}\delta (E-\zeta_{n})+|v_{n,\nu }|^{2}\delta
(E+\zeta _{n})],$ where $(u_{n,\nu},v_{n,\nu})^{T}$ is the
eigenvector corresponding to the eigenenergy $\zeta_{n}$ of the
mean-field BdG Hamiltonian with cylinder geometry. We find that the
{zero-energy} Majorana fermions manifest themselves by a peak in
LDOS located at the edges of the system, as shown in
Fig.~\ref{fig:tFFedge}(b). This spatially localized zero-energy peak
in LDOS can be detected using spatially resolved radio-frequency
(rf) spectroscopy technique~\cite{2007_Ketterle_PRL}, {which would
provide a concrete signature for the experiment.}

\begin{figure}[t]
\begin{center}
\includegraphics[width=9.5cm]{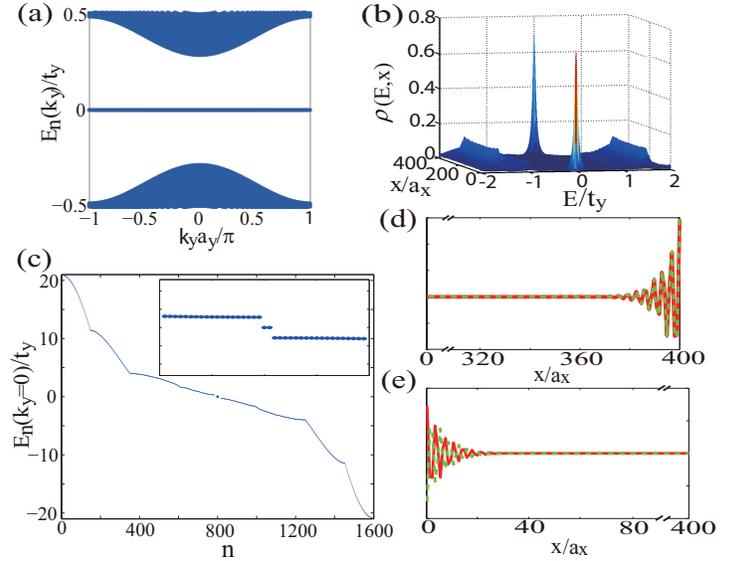}
\end{center}
\caption{{Topologically protected edge states composed of Majorana
fermions in the tFF superfluid state.} (a) Energy spectrum of the
Hamiltonian in Eq.~\eqref{Ham} under the mean-field approximation
with open (and periodic) boundary conditions in the $x$ ($y$)
directions. There is {a doubly degenerate flat band} composed of
Majorana fermions when the transverse tunneling is small. (b) The
local density of states (LDOS) {(see the main text).} The
zero-energy peaks of the LDOS are located at the two edges of the
system respectively. (c) {Energy spectra with $k_ya_y=0$. There are
two degenerate zero energy states.} (d) and (e), The wavefunctions
of the two Majorana zero-energy modes as shown in (c), which satisfy
$u_{0,p_x}=v^*_{0,p_x}$ at the right edge and
$u_{0,p_x}=-v^*_{0,p_x}$ at the left edge (other components of these
two wavefunctions, say $u_{0,p_y}$ and $v_{0,p_y}$, also satisfy the
same relations). These two states support two local Majorana
fermions at two outer edges of the system respectively. {Here we
choose finite length along the $x$ direction with lattice sites
$N=400$ and $\tilde{h}=0.7$, $t/t_y=0.6$. Other parameters are the
same as in Fig.~\ref{fig:Phsaediagram} and $n$ denotes the energy
level.}} \label{fig:tFFedge}
\end{figure}

We also consider the effect of small transverse tunneling on the
stability of tFF superfluids. The transverse hopping term restores
the 2D nature of the Fermi surface, which acquires a finite
curvature in the transverse direction. It suppresses the perfect
nesting, therefore it disfavors the tFF superfluid state. Our
numerical result indicates that the tFF superfluity remains stable
at small value of transverse tunneling. For example, the tFF
superfluid state, as shown in Fig.~\ref{fig:Phsaediagram}(a) with
$t/t_y=0.8$ and $\tilde{h}=0.7$, survives until $t^{\prime}_y/t_y$
reaches $0.2$ and the flat bands composed of Majorana fermions in a
cylinder geometry will be maintained~\cite{2013_chuanwei_arxiv}.
Beyond this value, the bulk gap will close and the tFF superfluids
become unstable.

We would like to {emphasize} that the topological non-trivial
properties in the noncentrosymmetric optical lattice arise directly
from symmetry-breaking induced orbital hybridization. This plays a
curial role in producing the topological FF superfluid state. {We
expect noncentrosymmetric optical lattices would provide a fertile
ground to support orbital hybridized topological phases in general.}

\textit{Conclusion.} {We have established a systematic approach to
control non-trivial orbital hybridization in a static
noncentrosymmetric optical lattice, which is shown to give rise to
unconventional topological properties. We find that this lattice
system in the presence of onsite rotation supports a novel
topological FF superfluid state when loaded with attractive
fermions. It features Majorana zero energy modes located at the
outer edges, leading to signatures in the local density of states as
a concrete experimental evidence of the topological superfluid
state.}

\textit{Acknowledgement.} This work is supported by AFOSR
(FA9550-12-1-0079), ARO (W911NF-11-1-0230), Overseas Collaboration
Program of NSF of China No. 11429402 sponsored by Peking University,
the Charles E. Kaufman Foundation, and The Pittsburgh Foundation (B.
L. and W. V. L.). X. L. is supported by LPS-MPO-CMTC, JQI-NSF-PFC and ARO-
Atomtronics-MURI.

\begin{widetext}

\begin{center}
{\Large\bf Supplementary Materials}
\end{center}

\renewcommand{\thesection}{S-\arabic{section}} \renewcommand{\theequation}{S%
\arabic{equation}} \setcounter{equation}{0} %  this will re-count eq from 1
\renewcommand{\thefigure}{S\arabic{figure}} \setcounter{figure}{0}

\section{BdG equation in momentum space}

Through introducing the superfluid pairing order parameter
$\Delta(\mathbf {r})=U\langle C_{p_y}(\mathbf {r})C_{p_x}(\mathbf
{r})\rangle=\Delta e^{i{\mathbf {Q}\cdot {\mathbf {r}}}}$, the
system can be described by the Bogoliubov-de-Genes (BdG) Hamiltonian
at mean field level
\begin{equation}
H_{BdG}(\mathbf {k})=\frac{1}{2}
\begin{pmatrix}
{\varepsilon_{p_x}(\mathbf {Q/2+k})} & {\varepsilon(\mathbf {Q/2+k})} & {0} & {\Delta} \\
{-\varepsilon(\mathbf {Q/2+k})} & {\varepsilon_{p_y}(\mathbf {Q/2+k})} & {-\Delta} & {0}\\
{0} & {-\Delta^{*}} & {-\varepsilon_{p_x}(\mathbf {Q/2-k})} & {\varepsilon(\mathbf {Q/2-k})} \\
{\Delta^{*}} & {0} & {-\varepsilon(\mathbf {Q/2-k})} &
{-\varepsilon_{p_y}(\mathbf {Q/2-k})}
\end{pmatrix} ,
\label{BdGHam}
\end{equation}
where the Nambu basis is chosen to be $(C_{p_x}(\mathbf {Q/2+k}),
C_{p_y}(\mathbf {Q/2+k}),$ $C_{p_x}^{\dagger}(\mathbf {Q/2-k}),
C_{p_y}^{\dagger }(\mathbf {Q/2-k}))^{T}$,
$\varepsilon_{p_x}(\mathbf {k})=2t_x \cos(k_xa_x)-2t^{\prime}_{x}
\cos(k_ya_y)-(\mu-h)$, $\varepsilon_{p_y}(\mathbf {k})=-2t_y
\cos(k_xa_x)+2t^{\prime}_{y} \cos(k_ya_y)-(\mu+h)$ and
$\varepsilon(\mathbf {k})=2it \sin(k_xa_x)+i\Omega_z$.

Then, the free energy can be obtained by diagonalizing the BdG
Hamiltonian in Eq.~\eqref{BdGHam} by standard procedure as
\begin{eqnarray*}
F[\Delta]&=& 1/2\sum_{\mathbf {k}}[\varepsilon_{p_x}(\mathbf
{k})+\varepsilon_{p_y}(\mathbf {k})
+\sum_{\lambda}\Theta(-E_{\lambda}(\mathbf {k}))E_{\lambda}(\mathbf
{k})]-\frac{N|\Delta|^2}{U}, \label{Freeen}
\end{eqnarray*}
where $E_{\lambda}$ is {the quasi-particle energy,} and $\Theta$ is
the Heaviside step function. The pairing order parameter $\Delta$
and the center-of-mass momentum of pairing $\mathbf {Q}$ can be
determined from minimizing the free energy.

\section{$Z_2$ topological invariant}
To characterize the topological nature of the tFF superfluid state,
we calculate the $Z_2$ topological invariant. Here, we introduce the
Majorana operators as $\gamma^A(\mathbf{r})=C_{p_x}^{\dagger
}(\mathbf{r})+C_{p_x}(\mathbf{r})$,
$\gamma^B(\mathbf{r})=[C_{p_x}(\mathbf{r})-C_{p_x}^{\dagger
}(\mathbf{r})]/i$, $\gamma^C(\mathbf{r})=C_{p_y}^{\dagger
}(\mathbf{r})+C_{p_y}(\mathbf{r})$ and
$\gamma^D(\mathbf{r})=[C_{p_y}(\mathbf{r})-C_{p_y}^{\dagger
}(\mathbf{r})]/i$, which fulfill the relations ${\gamma^{\dagger
}}^\alpha(\mathbf{r})=\gamma^\alpha(\mathbf{r})$ and the
anticommutation relations $\{ \gamma^\alpha,\gamma^\beta
\}=2\delta_{\alpha\beta}\delta(\mathbf{r-r'})$ with $\alpha$ or
$\beta$ taking A, B, C or D. In terms of Majorana operators, the
Hamiltonian in Eq. (4) under the mean-flied approximation can be
represented as
$H_{MF}=\frac{i}{4}\sum^{4N}_{l,m=1}A_{lm}\gamma_l\gamma_m$ with
$\gamma_{4j-3}=\gamma^A(\mathbf{r}_j)$,
$\gamma_{4j-2}=\gamma^B(\mathbf{r}_j)$,
$\gamma_{4j-1}=\gamma^C(\mathbf{r}_j)$ and
$\gamma_{4j}=\gamma^D(\mathbf{r}_j)$, where j runs over all the $N$
lattice sites and A is a skew-symmetric matrix. The Pfaffian of
matrix A is defined as $\mathrm{Pf}(A)=\frac{1}{2^nn!}\sum_{\tau \in
S_{2n}}sgn(\tau)\prod_{m=1}^{n}A_{\tau(2m-1),\tau(2m)}$ with $n=2N$,
where $S_{2n}$ is the set of permutation and $sgn(\tau)$ is the
corresponding sign of that. The $Z_2$ topological  invariant is
defined as $M=sgn[\mathrm{Pf}(A)]$ when choosing the periodic
boundary condition. The $Z_2$ topological non-trivial phase is
characterized by $M=-1$, where as the topological trivial phase
corresponds to $M=1$.

\end{widetext}

\bibliographystyle{apsrev}
\bibliography{tFF}

\end{document}